\documentclass[aps,prmaterials,groupedaddress,reprint,showpacs,dvipdfmx,floatfix]{revtex4-2}
\usepackage{graphicx}
\usepackage{comment}
\usepackage{tabularx}
\usepackage{dcolumn}
\usepackage{bm}
\usepackage{amsmath}
\usepackage{amssymb}
\usepackage{txfonts}
\usepackage{url}
\usepackage{xcolor}
\usepackage{multirow}
\usepackage{ulem}
\usepackage{subfigure}
\usepackage{siunitx}
\usepackage{here}
\usepackage[super]{nth}
\newcommand{\ghost}[1]{}

\makeatletter
\def\Hline{
\noalign{\ifnum0=`}\fi\hrule \@height 1pt \futurelet
\reserved@a\@xhline}
\makeatother

\begin{document}
\title{
  Diffusion Monte Carlo Study on Relative Stabilities of Boron Nitride Polymorphs
}
\author{Yutaka Nikaido$^{1}$}
%\email{mwk1805nkd@icloud.com}
\email{These two authors made the same contribution.}
\author{Tom Ichibha$^{2}$}
\email{These two authors made the same contribution.}
\author{Kenta Hongo$^{3}$}
\author{Fernando A. Reboredo$^{2}$}
\author{K.~C.~Hari~Kumar$^{4}$}
\author{Priya Mahadevan$^{5}$}
\author{Ryo Maezono$^{1}$}
\author{Kousuke Nakano$^{1,6}$}
\email{kousuke\_1123@icloud.com}

\affiliation{$^{1}$ School of Information Science, JAIST, Nomi, Ishikawa, 923-1292, Japan}
\affiliation{$^{2}$ Materials Science and Technology Division, Oak Ridge National Laboratory, Oak Ridge, Tennessee, 37831, USA}
\affiliation{$^{3}$ Research Center for Advanced Computing Infrastructure, JAIST, Asahidai 1-1, Nomi, Ishikawa, 923-1292, Japan}
%\affiliation{$^{4}$ Materials Science and Technology Division, Oak Ridge National Laboratory, Oak Ridge, Tennessee, 37831, USA}
\affiliation{$^{4}$ Department of Metallurgical and Materials Engineering, Indian Institute of Technology Madras, Chennai, 600036, India}
\affiliation{$^{5}$ S.N. Bose National Center for Basic Sciences, Block-JD, Salt lake, Kolkata, 700106, India}
\affiliation{$^{6}$ International School for Advanced Studies (SISSA), Via Bonomea 265, Trieste, 34136, Italy}
\date{\today}

\begin{abstract}  
Although Boron nitride (BN) is a well-known compound widely used for engineering and scientific purposes, the phase stability of its polymorphs, one of its most fundamental properties, is still under debate. The ab initio determination of the ground state of the BN polymorphs, such as hexagonal and zinc-blende, is 
difficult because of the elusive Van der Waals interaction, which plays a decisive role in some of the polymorphs, making quantitative prediction highly challenging. Hence, despite multiple theoretical studies, there has been no consensus on the ground state yet, primarily due to contradicting reports. In this study, we apply a state-of-the-art ab initio framework - fixed-node diffusion Monte Carlo (FNDMC), to four well known BN polymorphs, namely hexagonal, rhombohedral, wurtzite, and zinc-blende BNs. Our FNDMC calculations show that hBN is thermodynamically the most stable among the four polymorphs at 0 K as well as at 300K. This result agrees with the experimental data of Corrigan~{\it et al.} and Fukunaga. The conclusions are consistent with those obtained using other high-level methods, such as coupled cluster. We demonstrate that the FNDMC is a powerful method to address polymorphs that exhibit bonds of various forms. It also provides valuable information, like reliable reference energies, when reliable experimental data are missing or difficult to access. Our findings should promote the application of FNDMC for other van der Waals materials.

\end{abstract}
\maketitle

\section{Introduction}
\vspace{2mm}
Since most crystalline materials exist in only one type of structure, exceptions like polymorphs have sparked curiosity for centuries. These materials possess a delicate balance of bond formation energies and entropy. These occur due to the existence of multiple structures with the same chemical formula but strikingly different mechanical, optical, and electronic properties. Understanding how to control or favor a particular phase as a function of temperature and pressure is crucial for material design and highly relevant to planetary science. Accurate computational techniques have played important roles in the calculation of phase diagrams, particularly in extreme regions that are difficult to explore via experiment. For instance, in the particular case of Boron nitride (BN), four stable polymorphs – namely hexagonal, rhombohedral, wurtzite, and zinc-blende, coexist within a narrow energy window.
These are dubbed as hBN, rBN, wBN, and cBN in this paper, respectively. 
In addition, hBN can fold, forming nanotubes with different radii. 
Due to elastic strain, nanotube polymorphs of BN are expected to have higher energy than hBN~\cite{2003HJX_QZ}.
While there have been many experiments to determine the relative stabilities of the BN crystal polymorphs\cite{1963BUN, 1975COR, 1975TAN, 1987MIS, 1989KAG, 1993NAK, 1993SOL, 1995SOL, 1998ERE, 1999SOL, 2000FUK, 2000WIL, 2018WOL}, their relative phase stabilities are still under debate. We focus on the experiments reported by Solozhenko~{\it et al.}~{\cite{1993SOL, 1995SOL}}, Corrigan~{\it et al.}~{\cite{1975COR}}, and Fukunaga~{\cite{2000FUK}} because they are described the most probable phase stability at room temperature, and also determined the ground state at 0~K; therefore, they are comparable to our computational study. Solozhenko~{\it et al.} claimed that cBN is more stable than hBN under ambient conditions, while Corrigan~{\it et al.} and Fukunaga argued otherwise. Indeed, two conflicting results have been obtained, and it has been a long remaining question to clarify which of these experimental results is more accurate.
In such a case, accurate computational techniques could provide valuable clues to settle the debate.

\vspace{2mm}
Thus, there have been many attempts to discuss the stability of the BN polymorphs from a computational point of view~{\cite{1994FUR, 2003YU, 1997ALB, 1999KER, 1991YON, 2009TOPP, 1994FUR, 2018GRU, 2019CAZ, 2021HEL}}. Many studies on the DFT-LDA functional~{\cite{1994FUR, 2003YU, 1997ALB, 1999KER, 1991YON}} show that the cBN is more stable than hBN, which agrees with the experimental results by Solozhenko et al. However, the fact that hBN (and rBN) is a quasi-two-dimensional material makes it difficult to distinguish the relative stability of these polymorphs with standard mean-field ab initio calculations,
because the van der Waals interaction cannot be evaluated quantitatively withtin LDA and other semi-local approximations that neglect the non-locality of the exchange and correlation~{\cite{1998AND}}.
Instead, functionals that include semi-empirical vdW corrections (e.g., D3 correction) might be used to handle this problem within the DFT framework. However, the results are significantly dependent on the type of the vdW-correction, as shown later. 
Recently, more sophisticated methods, such as many-body dispersion (MBD)~\cite{2019CAZ}, random phase approximation (RPA)~\cite{2019CAZ}, RPA with exchange (RPAx)~\cite{2021HEL}, and coupled cluster with single, double, and perturbative triple excitations (CCSD(T))~\cite{2018GRU}, have been applied to study the BN stability problem. 
Nevertheless, the ground state of the BN polymorphs is still under debate, because they provided contradicting predictions of the relative stabilities.

\vspace{2mm}
Quantum Monte Carlo (QMC) methods~{\cite{2001FOU,2012BMA_WAL,2011JK_LM,2009WAL_BH,2020RJN_JRT,2020NAK,2009BOO,2010BOO,2010CLE,2011CLE,2004WP_SZ,2003SZ_HK}} are the state-of-the-art ab initio frameworks that do not lose quantitative predictability even for peculiar materials such as vdW solids.
There are two major steps in the most popular ab initio QMC approach: 1) variational Monte Carlo (VMC) and 2) fixed-node diffusion Monte Carlo (FNDMC). The former is a method that variationally optimizes a parameterized many-body wave function to reach the ground state (i.e., the variational principle). The latter obtains the ground state from a given trial function, optimized in VMC using the projection operator on the ground state without needing any heuristic parameterization. In particular, FNDMC is effective at quantitatively treating vdW forces~{\cite{2015MOS}} and has been applied to many peculiar compounds such as graphite~{\cite{2009SPA}}, cyclohexasilane dimers~\cite{2017KH_RMa}, DNA stacking~\cite{2015KH_RMa,2020KSQ_RMa}, black phosphorus~{\cite{2019FRA}},  boron nitride bilayer~{\cite{2014HSI}}, graphene bilayer~{\cite{2015MOS}}, molecular crystals~{\cite{2010KH_AAGa,2015KH_RMa,2018AZ_AMa}}, cyclodextrin-drug inclusion complexes~\cite{2020KO_RMa}, and hydrogen adsorption on nanotubes~\cite{2021GIP_KHa}, giving distinguishable results from DFT ones.

%%%%%%%%%%%%%%%%%%%%%%%%%%%%%%%%%%%%%%%%%
\begin{figure*}[htbp]
  \centering
  \includegraphics[width=\hsize]{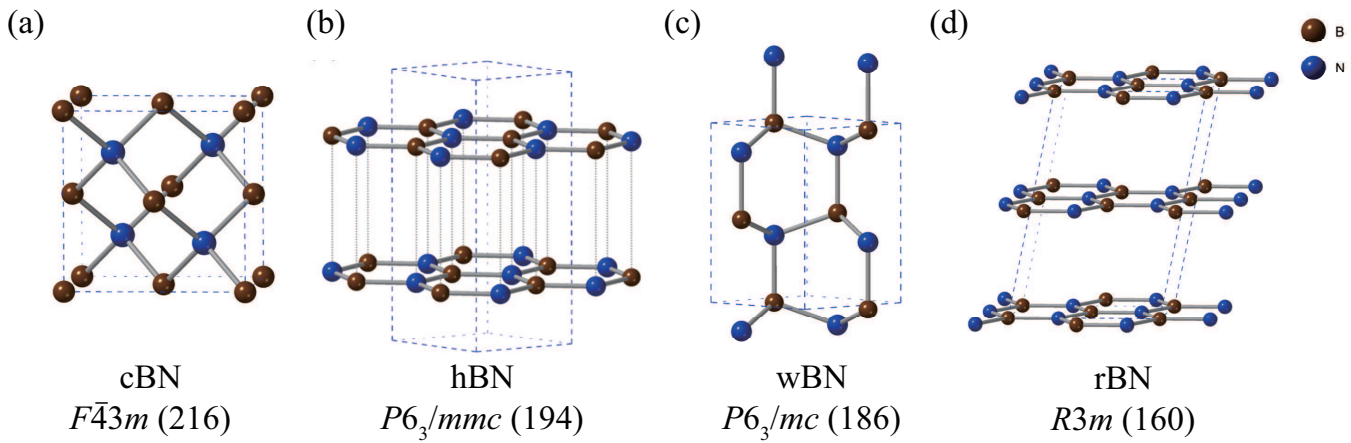}
  \caption{
    \label{fig.hbn}
    Structures of (a) cBN, (b) hBN, (c) wBN, and (d) rBN.
	Their space groups are also shown.
    Boron and Nitrogen atoms are colored in brown and blue, respectively.
  }      
\end{figure*}
%%%%%%%%%%%%%%%%%%%%%%%%%%%%%%%%%%%%%%%%%

\vspace{2mm}
In this study, we applied ab initio FNDMC calculations to the four crystal polymorphs of BN to discuss their relative phase stability. We found that hBN is the most thermodynamically stable among these four polymorphs, in agreement with the experiments done by Corrigan~{\it et al.} and Fukunaga.

%%%%%%%%%%%%%%%%%%%%%%%%%%%%%%%%%%%%%%%%%
\section{Theory of FNDMC}
\label{sec.theory}\ghost{sec.theory}
%%%%%%%%%%%%%%%%%%%%%%%%%%%%%%%%%%%%%%%%%
Here, we briefly explain the theory of FNDMC. 
Further details are found in several review papers on FNDMC 
~\cite{2001FOU, 2012BMA_WAL, 2011JK_LM, 2009WAL_BH, 2020RJN_JRT, 2020NAK}. 
FNDMC is based on the projection operation
%----------------------------------------
\begin{equation}
  \Phi \left( {\vec R,t} \right) = \exp \left( { - \hat Ht} \right)\Psi_T \left( {\vec R} \right).
  \label{eq.proj}
\end{equation}\ghost{eq.proj}
%----------------------------------------
Here,
$\Psi_T\left(\vec R\right)$ is the trial wave function,
$\hat H$ is the Hamiltonian, and $t$ is the simulation time. 
With increasing $t$, $\Phi \left( {\vec R,t} \right)$ gradually gets close to the ground state wave function,
because the projection term, $\exp \left( { - \hat Ht} \right)$, gets reduced faster for a higher energy eigen state.
The time-evolution of $\Phi \left( {\vec R,t} \right)$ is expressed as 
a kind of diffusion equation with the importance sampling technique~\cite{1971GRI,2001FOU}:
%----------------------------------------
\begin{eqnarray}
  \frac{{\partial f}}{{\partial t}} = &\frac{1}{2}&\sum\limits_i {\vec \nabla _i^2f}  - \sum\limits_i {{{\vec \nabla }_i} \cdot \left( {f{{\vec \nabla }_i}\ln {\Psi _T}} \right)} \nonumber \\
  &-& \left[ {\frac{{\left( {\hat H - {E_T}} \right)\Psi_T }}{\Psi_T }} \right]f, 
  \label{eq.impo1}
\end{eqnarray}\ghost{eq.impo1}
%----------------------------------------
where, 
%----------------------------------------
\begin{equation}  
    f\left( {\vec R,t} \right) = \Phi \left( {\vec R,t} \right){\Psi _T}\left( {\vec R} \right).
\end{equation}
%----------------------------------------
This diffusion equation is solved by the random walk algorithm with a variable number of walkers.

\vspace{2mm}
The most common trial wave function, $\Psi_T\left(\vec R\right)$, is the Slater-Jastrow type~\cite{1955JAS,2001FOU}:
%----------------------------------------
\begin{eqnarray}
  {\Psi _T}\left( {\vec R} \right) = \exp \left\{ {{J_{e - I}}\left( {\vec R} \right) + {J_{e - e}}\left( {\vec R} \right)}\right\}D^\uparrow\left(\vec R\right)D^\downarrow\left(\vec R\right)
  \label{eq.trial}
\end{eqnarray}\ghost{eq.trial}
%----------------------------------------
Here, ${{J_{e - I}}\left( {\vec R} \right)}$ and ${{J_{e - e}}\left( {\vec R} \right)}$ are
one- and two-body Jastrow factors.
They are non-negative parameterized functions that describe electron--ion and electron--electron
correlations, respectively.
The parameters are optimized by the ``linear'' method~\cite{2007CJU_RGH}.
$D^\sigma\left( {\vec R} \right)$ are Slater determinants for spin $\sigma$ generally obtained from DFT or Hartree--Fock method.
$f\left( {\vec R,t} \right)>0$ is imposed in FNDMC.
Thus, the nodal surfaces ($\Phi$=0) are fixed to those of ${\Psi _T}\left( {\vec R} \right)$.
This aims to make $\Phi \left( {\vec R,t} \right)$ anti-symmetric for parallel spin exchanges and  is called as fixed-node approximation.
Without this approximation, FNDMC would give a bosonic solution. 
An FNDMC result has a bias (fixed-node error) that depends on the quality of the nodal surfaces of the trial wave function~\cite{2001FOU}.

\vspace{2mm}
Semi-local potentials~\cite{2019GW_LM, 2018AA_LM, 2018MCB_LM, 2017MCB_LM, 2017JRT_RJN, 2016JTK_FAR, 2015JRT_RJN}
%----------------------------------------
\begin{equation}
  {\hat V_I}\left( r \right)=\hat V_I^{{\rm{local}}}\left( r \right) + \hat V_I^{{\rm{non-local}}}\left( r \right)
\end{equation}
%----------------------------------------
are generally used when the FNDMC is applied to solids.
With the semi-local potentials, Equation \eqref{eq.impo1} is written 
%----------------------------------------
\begin{eqnarray}
  \frac{{\partial f}}{{\partial t}} = &\frac{1}{2}&\sum\limits_i {\vec \nabla _i^2f}  - \sum\limits_i {{{\vec \nabla }_i} \cdot \left( {f{{\vec \nabla }_i}\ln {\Psi _T}} \right)} \nonumber \\
  &-& \left[ {\frac{{\left( {\hat H - {E_T}} \right)\Psi }}{\Psi }} \right]f - \varepsilon \left( {\vec R} \right)f,
  \label{eq.impo2}
\end{eqnarray}\ghost{eq.impo2}
%----------------------------------------
where,
%----------------------------------------
\begin{equation}
  \varepsilon \left( {\vec R} \right) = \left( {\frac{{\hat V_I^{{\rm{NL}}}\Phi }}{\Phi } - \frac{{\hat V_I^{{\rm{NL}}}{\Psi _T}}}{{{\Psi _T}}}} \right). \label{eq.locapp}
\end{equation} \ghost{eq.locapp}
%%----------------------------------------
However, ${\hat V_I^{{\rm{NL}}}\Phi }$ cannot be estimated because
FNDMC does not provide an analytical form of ${\Phi \left( {\vec R},t \right)}$.  
Therefore, locality approximation, $\varepsilon \left( {\vec R} \right)=0$~\cite{1991MIT}, is generally used.  
Under the approximation, Equation \eqref{eq.impo2} is identical to Equation \eqref{eq.impo1}.  
When ${{\Psi _T}}$ is the ground state wave function,  
$\varepsilon \left( {\vec R} \right)=0$ is proved~\cite{1991MIT} and 
the locality approximation is not needed. 
Therefore, the bias by the locality approximation (localization error)
tends to decrease by improvement in the quality of the trial wave function. 

\vspace{2mm}
With the locality approximation, the FNDMC energy is not variational
because the localization error can be both positive or negative~\cite{2006CAS}.
As an alternative method to utilize the semi-local potentials, the T-move approach~\cite{2006CAS},
selectively evaluated terms making positive biases in $\varepsilon \left( {\vec R} \right)$
with a heat bath algorithm~\cite{2006CAS}.
Therefore, the bias of the T-move approach is always positive, so the variational principle is maintained.
However, it is still debated as to which is the better method to deal with
$\varepsilon \left( {\vec R} \right)$ in practice~\cite{2019ZEN}.
A recently developed method, determinant localization approximation,
is claimed to be better than the locality approximation and T-move approach
in terms of the bias suppression, stability, and efficiency of calculation~\cite{2019ZEN}.
We used the T-move approach in this work since the determinant localization approximation is
not yet implemented in QMC code, QMCPACK~\cite{Kim2018}, that is used here.

%%%%%%%%%%%%%%%%%%%%%%%%%%%%%%%%%%%%%%%%%
\section{Computational details}
\label{sec.details}\ghost{sec.details}
%%%%%%%%%%%%%%%%%%%%%%%%%%%%%%%%%%%%%%%%%
%%%%%%%%%%%%%%%%%%%%%%%%%%%%%%%%%%%%%%%%%
\subsection{FNDMC calculation}
%%%%%%%%%%%%%%%%%%%%%%%%%%%%%%%%%%%%%%%%%
The FNDMC calculations were performed using QMCPACK~\cite{Kim2018} 
with Slater-Jastrow type trial wave function~\cite{2001FOU}, where
all the calculations were managed with a workflow management system, \textsc{Nexus} \cite{2016KRO}.
The orbitals of the Slater determinant were determined 
by DFT calculations with the vdW--DF2 functional~\cite{2010LEE}
using Quantum Espresso~\cite{Giannozzi2009}.
The plane-wave basis set with the energy cutoff equal to 350~Ry was employed,
with which the total energies of hBN, cBN, wBN, and rBN converged within 2.62 meV/f.u.
The correlation consistent effective core potentials (ECPs)~\cite{2017BEN} were employed
in this study for decreasing the computational cost. 
The Jastrow factor is composed of one and two body terms,
amounting 32 variational parameters in total.
The variational parameters were optimized by the so-called linear method \cite{2007UMR} implemented in QMCPACK.
We carefully estimated the controllable errors of FNDMC, i.e., 
the time step error and one- and two-body finite size errors~\cite{2014DRU}.
The timestep error was corrected by extrapolating the relative FNDMC total energies 
from hBN for time step, $dt$, towards the $dt=0$ limit.
Since the relative total energies were proportional to $dt$ below $dt = 0.04$~a.u.$^{-1}$
as shown in Fig.~\ref{fig.timestep}, we used $dt = 0.04$ and 0.01~a.u.$^{-1}$ for linear extrapolation.
The twist-averaging boundary condition was applied to eliminate the one-body finite size error. 
Figure \ref{fig.twist} shows the total energies of 16~f.u. simulation cell
calculated by the VMC method using QMCPACK for different twist grid sizes. Here, we prepared the simulation cells as isotropic as possible.
We employed the 2$\times$2$\times$2 grid for the FNDMC calculations,
since all the total energies converged within 1~kJ/mol with the grid size.
The two-body finite size errors were estimated by the size extrapolation:
the FNDMC relative total energies from hBN were extrapolated with respect to the inverse of the simulation cell size towards the infinite simulation cell size limit,
as shown in Figure \ref{fig.inverse}.
To achieve the smooth extrapolations, we used the Ewald method \cite{1921EWA} to evaluate the periodic potentials with the Gaussian charge screening breakup method.
  We chose this method rather than the default setting of the current version of QMCPACK (\textit{i.e.,} the optimized breakup method).
  This is because the optimized breakup method is known to cause a bias for some quasi-two-dimensional systems~\cite{GoogleGroup_1}.
  We carefully sufficiently reduced the error caused by excluding the nearest neighbor images in the real space portion of the Ewald term.
  The size of the real space portion is decided by the LR\_dim\_cutoff parameter in QMCPACK~\cite{Kim2018}.
  This quantity is defined as $\mathrm{LR\_dim\_cutoff} = r_{c} \times k_{c}$.
  Here, r$_c$ is the Wigner-Seitz radius and $k_c$ is the largest wave number
  to calculate the summations in the reciprocal space portion of the Ewald term.
A larger LR\_dim\_cutoff increases the evaluation accuracy of
the electron-electron periodic interactions. 
We confirmed that the error was less than 1~kJ/mol with LR\_dim\_cutoff equal to 20,
so we always used LR\_dim\_cutoff equal to or larger than 20.
%ECP
An uncontrollable error in our calculation is the so-called local error accompanied with the use of pseudopotentials (ECPs) at the FNDMC level.
We used the T-move approach~\cite{2006CAS} to alleviate the error as much as possible.
%geom
The experimental geometries taken from Ref.~{\onlinecite{1974SOM, 1993XU, 1952PEA}}
were used for cBN, wBN, and hBN, respectively, while that taken from Ref.~{\onlinecite{2018GRU}} was employed for rBN.
All the geometries are detailed in the appendix.

%%%%%%%%%%%%%%%%%%%%%%%%%%%%%%%%%%%%%%%%%%%%%%%%%%%%
\begin{center}
\begin{table*}[htbp]
  \caption{\ghost{Main\_comparison}\label{Main_comparison} Comparison of relative energies (in kJ/mol) of the four BN polymorphs (hBN, cBN, wBN, and rBN) with or without ZPE and TE, where ZPE and TE denote the zero-point energy and temperature effect, respectively. The energy of the most stable structure at 0~K among the four compounds is set as 0~kJ/mol for each method. Error bars in the CCSD(T) results are obtained from the fitting to estimate the finite size errors~\cite{2018GRU}.%\frcom{I would had written the table in the oposite order starting by FNDMC, CCSDT and so on. That would make more notorious the more accurate results. }
  }
  \vspace{2mm}
  \begin{tabular}{ccccccccc}
    \hline\hline
    \multirow{2}{*}{Method} & \multicolumn{4}{c}{without ZPE, without TE} & \multicolumn{4}{c}{with ZPE, with TE } \\
    \cline{2-9}
    & cBN & hBN & wBN & rBN & cBN & hBN & wBN & rBN\\
    %\hline
    %Exp.~1~{\footnotemark[1]} & - & - & - & - & 0.0 & 13.7 $\pm$ 3.0 & 3.0 $\pm$ 3.0 & 16.5 $\pm$ 4.1 \\ 
    %Exp.~2~{\footnotemark[2]} & 0.0 & 11.0 & - & - & 0.0 & 9.7 & - & - \\
    %Exp.~3~{\footnotemark[3]} & 0.7 & 0.0 & - & - & - & - & - & - \\
    %Exp.~4~ & - & - & - & - & - & - & - & - \\
    \hline
    DFT-LDA(PZ81)~{\footnotemark[3]} & 0.0 & 10.6 & 3.9 & - & - & - & - & - \\ 
    DFT-LDA(PZ81)~{\footnotemark[1]} & 0.0 & 14.1 & 3.3 & 13.8 & 0.0 & 11.9 & 4.8 & 11.8 \\
    DFT-GGA(PW91){\footnotemark[2]} & 15.4 & 0.0 & 19.3 & - & - & - & - & - \\ 
    DFT-GGA(PBE)~{\footnotemark[1]} & 13.3 & 0.0 & 17.1 & 1.2 & 15.5 & 0.0 & 20.7 & 1.3 \\ 
    DFT-vdW-DF2~{\footnotemark[1]} & 47.7 & 0.0 & 53.3 & 0.9 & 49.9 & 0.0 & 56.9 & 1.1 \\ 
    DFT-D3~{\footnotemark[5]} & 4.2 & 0.0 & 8.1 & 0.6 & - & - & - & - \\ 
    DFT-D3(BJ)~{\footnotemark[5]} & 0.0 & 0.5 & 3.6 & 0.9 & - & - & - & - \\ 
    MBD~{\footnotemark[5]} & 0.0 & 3.4 & 3.7 & 3.2 & - & - & - & - \\ 
	RPA~{\footnotemark[5]} & 0.0 & 2.5 & 3.7 & 2.8 & 0.0 & 0.3 & - & 0.8 \\ 
    RPAx~{\footnotemark[6]} & 1.4 & - & 5.0 & 0.0 & - & - & - & - \\ 
    CCSD(T)~{\footnotemark[4]} & 2.3(2.1) & - & 6.9(2.9) & 0.0 & - & - & - & - \\ 
    FNDMC~{\footnotemark[1]} & 3.5(4) & 0.0 &  7.8(4) & 1.5(4) & 5.6(4) & 0.0 & 11.5(4) & 1.7(4) \\
	\hline\hline
  \end{tabular}
  %
  %\footnotetext[1]{Solozhenko~{\it et al}~\cite{1995SOL} at T$=$298.15K under standard atmosphere.}
  %\footnotetext[2]{Jeong~{\it et al}~\cite{2013JEO} at T = 0 K without ZPE as extrapolated by Gruber~{\it et al}.~\cite{2018GRU}}
  %\footnotetext[3]{Day~\cite{2012DAY} at T = 0 K without ZPE as extrapolated by Gruber~{\it et al}.~\cite{2018GRU}}
  \footnotetext[1]{This work.}
  \footnotetext[2]{Topsakal~{\it et al}~\cite{2009TOPP}.}
  \footnotetext[3]{Furthm\"uller~{\it et al}~\cite{1994FUR}.}
  \footnotetext[4]{Gruber~{\it et al}~\cite{2018GRU}.}
  \footnotetext[5]{Cazorla~{\it et al}~\cite{2019CAZ}.}
  \footnotetext[6]{Hellgren and Bagunet~\cite{2021HEL}.}
\end{table*}
\end{center}
%%%%%%%%%%%%%%%%%%%%%%%%%%%%%%%%%%%%%%%%%%%%%%%%%%%%

%%%%%%%%%%%%%%%%%%%%%%%%%%%%%%%%%%%%%%%%%
\subsection{DFT and Phonon calculation}
%%%%%%%%%%%%%%%%%%%%%%%%%%%%%%%%%%%%%%%%%
The ab initio calculations based on the DFT were performed to compare energies
and compute vibrational and thermodynamical properties of BN polymorphs.
We used the Quantum Espresso~\cite{Giannozzi2009} package for these calculations.
The same norm-conserving PPs as used in the FNDMC calculations were employed
in the self-consistent field calculations.
These calculations were performed for the functionals of
LDA~\cite{1981PER}, PBE~\cite{1996PER}, and vdW-DF2~\cite{2010LEE}.
Cut off energy was set to 350 Ry. 
$k$-mesh grids were set to 7$\times$7$\times$3, 4$\times$4$\times$4, 6$\times$6$\times$4,
and 7$\times$7$\times$5 for hBN, cBN, wBN, and rBN, respectively.
The total energies of the polymorphs converged within 2.66 meV/f.u
with the cut off energy and $k$-mesh grids.

\vspace{2mm}
The ab initio phonon calculations were performed using the finite-displacement method 
implemented in PHONOPY package \cite{2008TOG, 2015TOG} to assess the vibrational stability and to estimate 
their Gibbs energy as a function of temperature and pressure 
within the quasi-harmonic approximation (QHA) framework.
Since the phonon dispersion calculation at the QMC level is computationally very demanding 
at present, we estimated the FNDMC (DFT) Gibbs energy,
$G_{\mathrm{FNDMC\left(DFT\right)}}\left( p, T \right)$, from the LDA Gibbs energy,
$G_{LDA}\left( p, T \right)$, as done in the previous studies~{\cite{2018GRU, 2019CAZ}}.
%----------------------------------------
\begin{equation}
\label{aprxEq}
G_{\mathrm{FNDMC\left(DFT\right)}} \left( p, T \right) = E_{\mathrm{FNDMC\left(DFT\right)}} + \left(G_{\mathrm{LDA}} \left( p, T \right) - E_{\mathrm{LDA}}\right),
\end{equation}
%----------------------------------------
where, $E_{\mathrm{FNDMC\left(DFT\right)}}$ and $E_{\mathrm{LDA}}$ are the FNDMC/DFT and LDA energies at $p=0$~GPa and $T=0$~K.
Indeed, we always estimated the vibrational contributions with the LDA functional. This is reasonable because 
Cazorla~{\it et al.}~\cite{2019CAZ} showed that ZPE differences between BN polymorphs are
practically insignificantly dependent on the functional.
The Gibbs energy was obtained by minimizing Helmholtz energy $F \left( T, V \right)$ with $pV$ term for volume,
%----------------------------------------
\begin{equation}
\label{GibbsEq}
G \left( p, T \right) = \underset{V}{min} \left[ F\left( T, V \right) + pV \right],
\end{equation}
%----------------------------------------
where Helmholtz vibrational energy reads:
%----------------------------------------
\begin{equation}
\label{HelmholtzEq}
F \left( T, V \right) = E_{\mathrm{0}} + k_B T \sum_{\mathrm{q}} \sum_{\mathrm{j}} \ln \left\{ 2 \sinh \left[ \frac{\hbar \omega _{\mathrm{j}} \left( q, V \right)}{2 k_B T} \right] \right\}.
\end{equation}
%----------------------------------------
The force constant matrices used for the phonon calculations were obtained with
4$\times$4$\times$4, 8$\times$8$\times$4, 8$\times$8$\times$4, 6$\times$6$\times$4 
conventional cells for cBN (512 atoms), hBN (1024 atoms), rBN (512 atoms), and wBN (576 atoms), respectively.
The ultra-soft PPs were used for the phonon calculations with the LDA exchange-correlation functional.
Cut off energies were 50 Ry for hBN, rBN, and hBN, and 100 Ry for cBN, respectively.
$k$-mesh grid-sizes were 1$\times$1$\times$1 for all of the polymorphs.
In the PHONOPY's implementation, $ F \left( T, V \right)$ is fitted by Vinet's equation of state (EoS)~\cite{1989VIN}.
22, 22, 17, and 17 structures with different volumes were used for hBN, rBN, wBN, and cBN, respectively, when fitting the EoS to obtain the Helmholtz vibrational energy. Once we obtained the Gibbs energies for the polymorphs, $p$-$T$ phase-diagrams were also obtained straightforwardly.

%%%%%%%%%%%%%%%%%%%%%%%%%%%%%%%%%%%%%%%%%%%
\begin{figure}[htbp]
  \centering
  \includegraphics[width=\hsize]{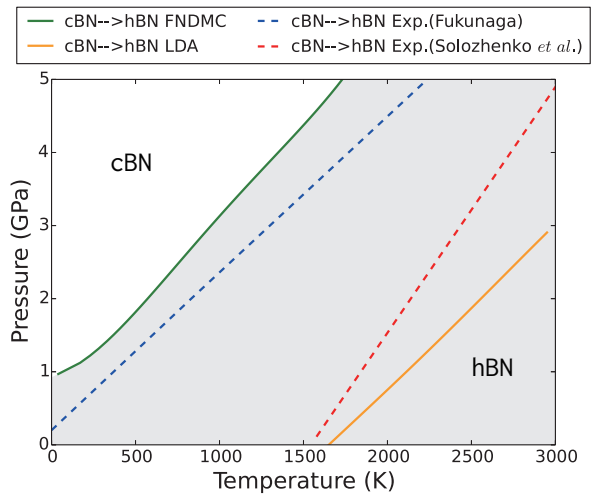}
  \caption{
    \label{fig.hzdiagram}
    (Color online) Equilibrium phase boundary between hBN and cBN.
The calculated FNDMC (green) and LDA (orange) equilibrium phase boundaries are 
depicted by solid lines.
Gray region is the cBN phase predicted by FNDMC. 
Note that error bars in FNDMC calculation are so small that they almost coincide with the FNDMC solid line.
Experimental data (Solozhenko~{\it et al.}~{\cite{1995SOL}} (red) and Fukunaga~{\cite{2000FUK}}) (blue) are also shown
by dotted lines.
  }      
\end{figure}
%%%%%%%%%%%%%%%%%%%%%%%%%%%%%%%%%%%%%%%%%%%%

%%%%%%%%%%%%%%%%%%%%%%%%%%%%%%%%%%%%%%%%%%%
\section{Results and discussion}
\label{sec.results}
%%%%%%%%%%%%%%%%%%%%%%%%%%%%%%%%%%%%%%%%%%%
\vspace{2mm}
Table~{\ref{Main_comparison}} shows comparisons of relative energies of  four BN representative polymorphs (hBN, cBN, wBN, and rBN)
With or without ZPE and TE. 
The choice of the exchange-correlation functionals has a significant effect on the evaluation of relative stability.
Indeed, the DFT calculations with the LDA (PZ81) suggest that cBN is the most stable structure, which agrees with the experiment
done by Solozhenko~{\it et al.}~\cite{1993SOL}.
On the other hand, the DFT calculations with the GGA (PW91 and PBE) suggest that hBN is the most stable structure, which 
agrees with the experiments done by Corrigan~{\it et al.}~\cite{1975COR} and Fukunaga~{\cite{2000FUK}}.
The DFT calculations with the vDW corrections (DF2 and DF3) also suggest that hBN is the most stable structure, while that with the Becke-Johnson damping (D3(BJ)) suggests that cBN is slightly more stable than hBN. 
Our result shows that the relative stabilities of the BN polymorphs are strongly dependent on the choice of exchange correlational functional in the DFT calculation. This is why a quantitative conclusion could not be made based on DFT results and need a more accurate method.

\vspace{2mm}
%\frcom{I would move this content to the beginning of the after the first centence describing the figure. The current placement makes it  difficult to find to a quick reader. Further, one often discusses new results first. DFT functionals comes later}
Our FNDMC result suggests that hBN is the most stable structure and the relative stability of the polymorphs is
hBN $\sim$ rBN $<$ cBN $<$ wBN at 0~K as well as at 300~K. The FNDMC calculation agrees the experiments done by Corrigan~{\it et al.}~\cite{1975COR} and Fukunaga~{\cite{2000FUK}}. The relative stability is consistent with the recent results obtained by CCSD(T)~{\cite{2018GRU}} and Random phase approximation with exchange (RPAx)~{\cite{2021HEL}} calculations and contradicts the recent work of Cazorla and Gould~{\cite{2019CAZ}}, claiming that cBN is the most thermodynamically stable among the four polymorphs, based on their RPA calculations. The quantitative reliabilities of these results are discussed later.
Fig.~\ref{fig.hzdiagram} shows the $p$-$T$ phase diagram of the cBN and hBN.
The phase diagram also suggests that hBN is more stable than cBN below 1.4~GPa at 0~K, which agrees with Fukunaga's experimental result \cite{2000FUK} rather than Solozhenko's one~\cite{1993SOL}.

%General
\vspace{2mm}
As already mentioned in the Introduction section, many computational works have been carried out to reveal the ground state of the BN polymorphs. For instance, DFT-LDA~\cite{1994FUR}, D3-BJ~\cite{2019CAZ}, MBD~\cite{2019CAZ}, and RPA~\cite{2019CAZ} calculations 
yield the cBN is the ground state, while DFT-PBE (this work), DFT-GGA(PW91)~\cite{2009TOPP}, RPAx~{\cite{2021HEL}}, CCSD(T)~\cite{2018GRU}, 
and FNDMC (this work) find that hBN (rBN)is the ground state. Thus, one would wonder which results are reliable.

\vspace{2mm}
As expected, simple DFT functionals such as PZ(LDA) and PBE(GGA) are not suitable for evaluating vdW interactions.
The dispersion-corrected DFT (DFT-D) is a simple but very powerful solution to capture vdW interactions within the DFT framework
by incorporating empirical potentials inversely proportional to \nth{6} of pairwise distance, 
proposed by Grimme et al.~{\cite{2016GRI}}.
The D3(BJ) method~{\cite{2010GRI, 2011GRI}} is a recent improvement of the older D2 approach~{\cite{2006GRI}}
using the Beck-Johnson dumping function~{\cite{2005BEC, 2005JOH, 2006JOH}} that determines the short-range behavior of the dispersion correction and is needed to avoid near-singularities for small pairwise distance region~{\cite{2011GRI}}.
In 2018, Tawfik et al. performed a benchmark test for 11 different DFT-vDW functionals, including D3(BJ)~{\cite{2018TAW}}, where the lattice constants were compared to the experimental values, while the binding energies were compared to the RPA ones. The results are divided into two groups: (1) the lattice parameters were well reproduced, but the binding energy was biased; on the other hand, (2) the binding energy was reproduced well, but the lattice parameters were biased. The D3(BJ) results belong to the first group. D3(BJ) shows rather large errors in the evaluation of the binding energy of layered materials, implying that the relative energy evaluation of the 2D material such as hBN is not quantitatively reliable. Instead, Table~{\ref{Main_comparison}} shows that the simple DFT-D3 gives closer results to the DMC values for the BN polymorphs.

%MBD
\vspace{2mm}
MBD method proposed by Tkatchenko et al.~{\cite{2012TKA}} provides a more general and less empirical approach that goes beyond the pairwise-additive treatment. 
It was reported that the MBD approach outperforms other vdW methods even in small systems~{\cite{2016FOR}}.
However, the binding energy evaluation is not satisfactory compared with CCSD(T) that undoubtedly provides highly accurate binding energies for molecules. For instance, it was reported that mean absolute error (MAE) of the binding energies between MBD and CCSD(T) $\sim$ 16.7~kJ/mol~{\cite{2017HER}} for the S12L set include relatively large molecules where vdW interactions mostly dominate the total binding energies. From the perspective of the order of $\sim$ 1~kJ/mol difference in the BN polymorphs, MBD does not seem accurate enough.

%RPA
\vspace{2mm}
DFT-RPA method is based on the RPA approximation that captures electron correlations which are not considered by the usual DFT functional. DFT-RPA has been benchmarked extensively for 2D layered materials{\cite{2010LEB, 2012BJO, 2017LEC}}. RPA results are often taken as ``reference values for exact solutions" for 2D layered materials~{\cite{2018TAW}}. 
However, it has been argued that RPA has a major problem since ``self-interactions" are not completely canceled out~{\cite{2009HAR}}. Therefore, in practice, the RPA function is known to sometimes underestimates the vDW interaction energies by approximately 20\% ~{\cite{2010ZHU, 2017DIX, 2018HEL}}, which is somewhat detrimental to the quantitative evaluation of the binding energy of 2D layered materials. In fact, RPA does not always work well and frequently underestimates the experimental values for compounds with dominant vDW forces~{\cite{2012REN, 2018ZEN}}. Furthermore, the conclusion drawn by Cazorla and Gould~\cite{2019CAZ} (i.e., cBN is the ground state) was obtained using RPA, but RPAx, an improved version of RPA, obtains that rBN as the ground state~{\cite{2021HEL}}.

%RPAx
\vspace{2mm}
RPAx was proposed to compensate for the shortcomings (i.e., self-interaction problems) of the RPA~{\cite{1998GOR, 2008HEL, 2010HEL, 2010HES}}. The RPAx goes beyond the ring diagrams to incorporate higher-order exchange effects and thus is known to reduce the self-interaction errors~{\cite{2014COL}}. Recently, RPAx was applied to polymorphs of BNs, and the relative energies of cBN, wBN, and rBN~{\cite{2021HEL}}, where RPAx gives a different result from RPA, i.e., rBN is more stable than cBN. Since RPAx improves on the shortcomings of RPA~{\cite{2021HEL}}, the results given by RPAx should be more reliable than those of RPA. However, Ref.~{\onlinecite{2021HEL}} does not consider the relative stability between hBN and other polymorphs.

%CCSD(T)
\vspace{2mm}
%\frcom{This paragraph can be reorganized since it repeats ideas}
CCSD(T) and FNDMC are often used to provide the reference energy values for benchmark studies.
Generally, CCSD(T) is used for small molecules and FNDMC is used for larger molecules and solids.
The calculation cost of CCSD(T) increases rapidly with system size, scaling as $N^7$.
In addition, a very large basis set is needed
to estimate the error dependence on the basis set size~\cite{1996MAR, 1998HAL, 1998TRU}.
On the other hand, FNDMC needs more human efforts regardless of the system size.
CCSD(T) and FNDMC agree with each other for molecules at equilibrium geometries~\cite{2021ALH}. 
Yet in some peculiar cases, they give different results. 
For example, CCSD(T) sometimes fails for systems away from its equilibrium geometries~\cite{2008AGT_RJB},
while FNDMC does not have such a weakness.
Although CCSD(T) is considered among the most accurate methods for molecular systems,
its reliability for bulk materials has not been fully investigated.
On the other hand, FNDMC has some systematic errors as explained in Section \ref{sec.theory}.
After all, it is difficult to conclude which of FNDMC or CCSD(T) would be more accurate for the bulk BNs, 
because both methods include totally different systematic errors. 
Table \ref{Main_comparison} shows that FNDMC and CCSD(T) calculations with large enough basis sets agree with each other quantitatively: 
they both find the layered BNs, hBN and rBN, to be low energy. 
This implies that the different systematic errors in FNDMC and CCSD(T) may be both small.
Therefore, we conclude that the layered BNs are likely the most stable structures
at zero temperature and zero pressure. 
  
%%%%%%%%%%%%%%%%%%%%%%%%%%%%%%%%%%%%%%%%%
\section{conclusion}
\label{sec.conc}\ghost{[sec.conc]}
%%%%%%%%%%%%%%%%%%%%%%%%%%%%%%%%%%%%%%%%%
In this study, we applied the ab initio FNDMC method to four BN polymorphs,
hBN, rBN, wBN, and cBN, to investigate their relative stability. 
The Gibbs energy of BN polymorphs at finite temperature were estimated by DFT phonon calculations in addition to the FNDMC energy 
We found that hBN is the most thermodynamically stable at ambient pressure among the four crystalline polymorphs at 0~K as well as 300~K, which is consistent with
the experiments performed by Corrigan~{\it et al.} and Fukunaga. So far, many works studying the relative stability of the BN polymorphs have been published with various ab initio frameworks such as DFT (LDA and GGA), DFT-D, MBD, CCSD(T), RPA, and RPAx. CCSD(T), FNDMC, and RPAx, which are the state-of-the-art methods in different frameworks, all support that hBN (rBN) is the ground state, i.e., that the ground state of BN is hBN (rBN) and not cBN. Therefore, we conclude that the ab initio calculations support the experimental results of Corrigan~{\it et al.}~{\cite{1975COR}}, and Fukunaga~{\cite{2000FUK}} (i.e., hBN is the ground state and cBN is a metastable phase). Our finding showcases the ability of FNDMC for discerning subtle energy differences involving vdW forces and encourage
the application of the FNDMC for other van der Waals materials.

%%%%%%%%%%%%%%%%%%%%%%%%%%%%%%%%%%%%%%%%%
\section*{Acknowledgments}
%%%%%%%%%%%%%%%%%%%%%%%%%%%%%%%%%%%%%%%%%
%JAIST-kagayaki
The computations in this work have been mainly performed with computational resources 
from the facilities of Research Center for Advanced Computing Infrastructure at 
Japan Advanced Institute of Science and Technology (JAIST).
%Oakridge
T.I. acknowledges computational resources provided by the Oak Ridge Leadership
Computing Facility at Oak Ridge National Laboratory,
which is a user facility of the Office of Science of the US Department of Energy
under Contract No. DE-AC05-00OR22725, and by the Compute and Data Environment
for Science (CADES) at Oak Ridge National Laboratory.
% K.N. financial support.
K.N. acknowledges a support from the JSPS Overseas Research Fellowships, that from Grant-in-Aid for Early-Career Scientists Grant Number JP21K17752, and that from Grant-in-Aid for Scientific Research(C) Grant Number JP21K03400.
% T.I. financial support.
T.I.
and F.A.R.
supported by the US Department of Energy, Office of Science,
Basic Energy Sciences, Materials Sciences and Engineering Division.
% K.H. financial support.
K.H. is grateful for financial support from 
the HPCI System Research Project (Project ID: hp210019, hp210131, and jh210045),
MEXT-KAKENHI (JP16H06439, JP17K17762, JP19K05029, JP19H05169, and JP21K03400),
and the Air Force Office of Scientific Research
(Award Numbers: FA2386-20-1-4036).
% R.M. financial support.
R.M. is grateful for financial supports from 
MEXT-KAKENHI (JP16KK0097, JP19H04692, and JP21K03400), 
FLAGSHIP2020 (project nos. hp190169 and hp190167 at K-computer), 
the Air Force Office of Scientific Research 
(AFOSR-AOARD/FA2386-17-1-4049;FA2386-19-1-4015), 
and JSPS Bilateral Joint Projects (with India DST). 

\clearpage

%%%%%%%%%%%%%%%%%%%%%%%%%%%%%%%%%%%%%%%%%
\appendix
%%%%%%%%%%%%%%%%%%%%%%%%%%%%%%%%%%%%%%%%%

\makeatletter
\renewcommand{\refname}{}
\renewcommand*{\citenumfont}[1]{#1}
\renewcommand*{\bibnumfmt}[1]{[#1]}
\makeatother

%%%%%%%%%%%%%%%%%%%%%%%%%%%%%%%%%%%%%%%%%
\section{Preliminaries for energetics}
%%%%%%%%%%%%%%%%%%%%%%%%%%%%%%%%%%%%%%%%%

\setcounter{table}{0}
\setcounter{equation}{0}
\setcounter{figure}{0}
\renewcommand{\thetable}{A-\Roman{table}}
\renewcommand{\thefigure}{A-\arabic{figure}}
\renewcommand{\theequation}{A-\arabic{equation}}

\begin{figure}[htbp]
  \centering
  \includegraphics[width=1.0\hsize]{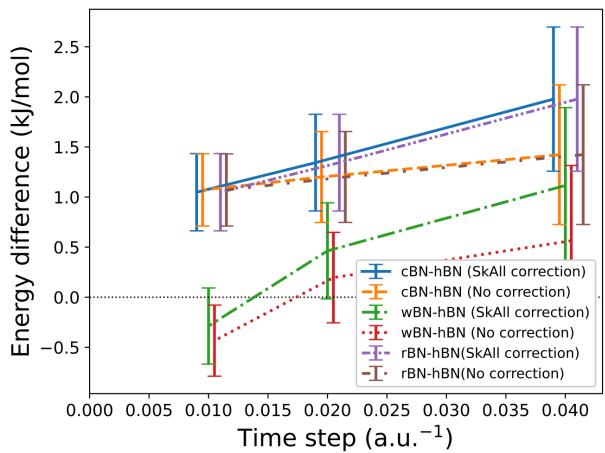}
  \caption{
    \label{fig.timestep}\ghost{[fig.timestep]}    
    Total energy differences per mole of formula unit predicted by FNDMC 
    with different time steps: 0.04, 0.02, and 0.01 a.u.$^{-1}$.
    Every data-series is slightly shifted in $x$ direction
    for the convenience of visualization.
    The given simulation cell consists of 16 f.u. and
    the twist-averaging bound condition was not used. 
  }      
\end{figure}

\begin{figure}[htbp]
  \centering
  \includegraphics[width=1.0\hsize]{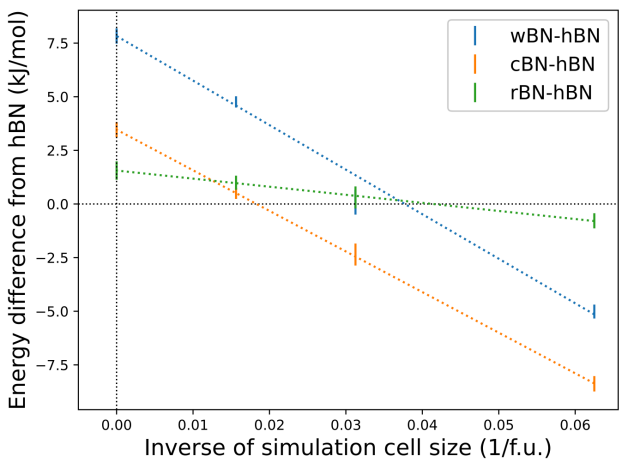}
  \caption{
    \label{fig.inverse}\ghost{[fig.inverse]}    
    Total energy differences per mole of formula unit predicted by FNDMC 
    for the inverse of different simulation cell sizes.
    The y-intercept value indicates the total energy
    excluding the two-body finite size error. 
  }      
\end{figure}

\begin{figure}[htbp]
  \centering
  \includegraphics[width=1.0\hsize]{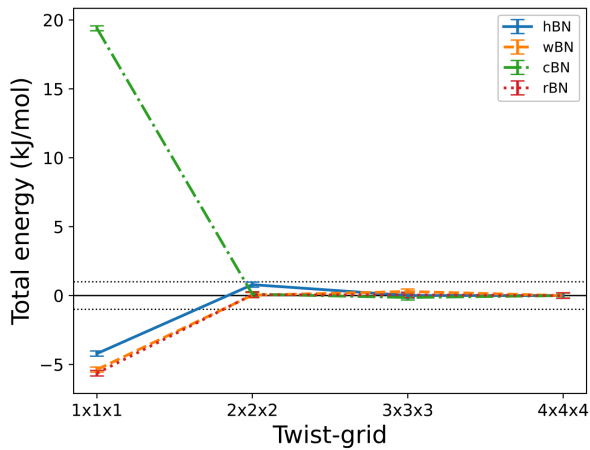}
  \caption{
    \label{fig.twist}\ghost{[fig.twist]}    
    Total energies per mole of formula unit of the monoclinic structure predicted
    by VMC with different grid sizes of twist-averaging.
    The total energies are given as relative differences
    from that with 4$\times$4$\times$4 grid.
    The dotted horizontal lines indicate 1 kJ/mol.
    The given simulation cell consists of 16 f.u..
  }      
\end{figure}

\begin{figure}[htbp]
  \centering
  \includegraphics[width=1.0\hsize]{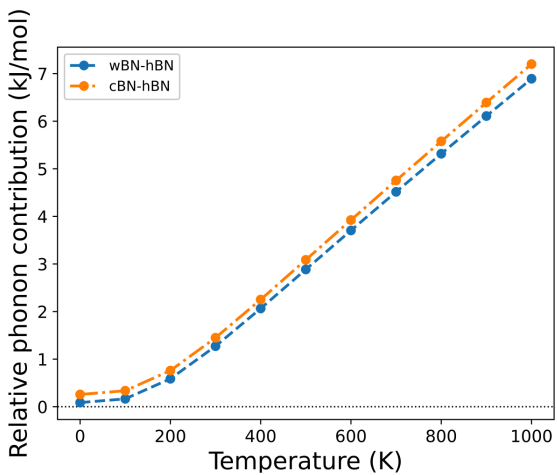}
  \caption{
    \label{fig.free}\ghost{[fig.free]}
    Differences of free energy of phonons per mole of formula unit
    between wBN (cBN) and hBN.
  }      
\end{figure}

We hereby summarize energetics calculated in this work,
which include energy dependency on the time steps (FIG.~{\ref{fig.timestep}}),
simulation cell size (FIG.~{\ref{fig.inverse}}), and
twist-grid (FIG.~{\ref{fig.twist}}).
Differences of free energy of phonons 
between polymorphs are also shown (FIG.~{\ref{fig.free}}).

%----------------------------------------
%----------------------------------------

%%%%%%%%%%%%%%%%%%%%%%%%%%%%%%%%%%%%%%%%%%%
%%%%%%%%%%%%%%%%%%%%%%%%%%%%%%%%%%%%%%%%%%%

%%%%%%%%%%%%%%%%%%%%%%%%%%%%%%%%%%%%%%%%%%%
%%%%%%%%%%%%%%%%%%%%%%%%%%%%%%%%%%%%%%%%%%%

%%%%%%%%%%%%%%%%%%%%%%%%%%%%%%%%%%%%%%%%%%%
%%%%%%%%%%%%%%%%%%%%%%%%%%%%%%%%%%%%%%%%%%%

%\clearpage
\newpage

%%%%%%%%%%%%%%%%%%%%%%%%%%%%%%%%%%%%%%%%%
\section{Geometries}
%%%%%%%%%%%%%%%%%%%%%%%%%%%%%%%%%%%%%%%%%

\setcounter{table}{0}
\setcounter{equation}{0}
\setcounter{figure}{0}
\renewcommand{\thetable}{B-\Roman{table}}
\renewcommand{\thefigure}{B-\arabic{figure}}
\renewcommand{\theequation}{B-\arabic{equation}}

We hereby summarize geometries used in this work.

%%%%%%%%%%%%%%%%%%%%%%%%%%%%%%%%%%%%%%%%%%%%%%%%%%%%
\begin{table*}[htbp]
  \caption{
Experimental structural parameters of wBN taken from Ref.~{\onlinecite{1993XU}}.
The lattice parameters are 
$a$ = 2.536000 \AA, 
$b$ = 2.536000 \AA, and
$c$ = 4.199000 \AA.
The angles are
$\alpha$ = \ang{90.0},
$\beta$ = \ang{90.0}, and
$\gamma$ = \ang{120.0}.
The space group is $P6_3mc$ (186).
}
\begin{center}
  \begin{tabular}{cccccc}
\hline\hline
Label & Wyckoff & $x$ & $y$ & $z$ \\
\hline
B &  2$b$  &  0.6667  &  0.3333  &  0.5000  \\
N &  2$b$  &  0.6667  &  0.3333  &  0.1250  \\
\hline\hline
\end{tabular}
\end{center}
\end{table*}
%%%%%%%%%%%%%%%%%%%%%%%%%%%%%%%%%%%%%%%%%%%%%%%%%%%%

%%%%%%%%%%%%%%%%%%%%%%%%%%%%%%%%%%%%%%%%%%%%%%%%%%%%
\begin{table*}[htbp]
  \caption{
Experimental structural parameters of cBN 
taken from Ref.~{\onlinecite{1974SOM}}.
The lattice parameters are 
$a$ = 3.615700 \AA, 
$b$ = 3.615700 \AA, and
$c$ = 3.615700 \AA.
The angles are
$\alpha$ = \ang{90.0},
$\beta$ = \ang{90.0}, and
$\gamma$ = \ang{90.0}.
The space group is $F\bar{4}3m$ (216).
}
\begin{center}
  \begin{tabular}{ccccc}
\hline\hline
Label & Wyckoff & $x$ & $y$ & $z$ \\
\hline
B & 4$a$ & 0.000000  &  0.000000  &  0.000000  \\
N & 4$c$ & 0.250000  &  0.250000  &  0.250000  \\
\hline\hline
\end{tabular}
\end{center}
\end{table*}
%%%%%%%%%%%%%%%%%%%%%%%%%%%%%%%%%%%%%%%%%%%%%%%%%%%%

%%%%%%%%%%%%%%%%%%%%%%%%%%%%%%%%%%%%%%%%%%%%%%%%%%%%
\begin{table*}[htbp]
  \caption{
Experimental structural parameters of hBN
taken from Ref.~{\onlinecite{1952PEA}}.
The lattice parameters are 
$a$ = 2.504000 \AA, 
$b$ = 2.504000 \AA, and
$c$ = 6.661200 \AA.
The angles are
$\alpha$ = \ang{90.0},
$\beta$ = \ang{90.0}, and
$\gamma$ = \ang{120.0}.
The space group is $P6_3/mmc$ (194).
}
\begin{center}
  \begin{tabular}{ccccc}
\hline\hline
Label & Wyckoff & $x$ & $y$ & $z$  \\
\hline
B & 2$c$ & 0.333333  &  0.666667  &  0.250000  \\
N & 2$d$ & 0.333333  &  0.666667  &  0.750000  \\
\hline\hline
\end{tabular}
\end{center}
\end{table*}
%%%%%%%%%%%%%%%%%%%%%%%%%%%%%%%%%%%%%%%%%%%%%%%%%%%%

%%%%%%%%%%%%%%%%%%%%%%%%%%%%%%%%%%%%%%%%%%%%%%%%%%%%
\begin{table*}[htbp]
  \caption{
Optimized structural parameters of rBN
taken from Ref.~{\onlinecite{2018GRU}}
The lattice parameters are 
$a$ = 2.487780 \AA, 
$b$ = 2.487780 \AA, and
$c$ = 3.534294 \AA.
The angles are
$\alpha$ = \ang{69.393402},
$\beta$ = \ang{69.393501}, and
$\gamma$ = \ang{60.000000}.
The space group is $R3m$ (160).
}
\begin{center}
  \begin{tabular}{ccccc}
\hline\hline
Label & Wyckoff & $x$ & $y$ & $z$  \\
\hline
B & 1$a$ & 0.000000  &  0.000000  &  0.000000  \\
N & 1$a$ & 0.333334  &  0.333333  &  0.000000  \\
\hline\hline
\end{tabular}
\end{center}
\end{table*}
%%%%%%%%%%%%%%%%%%%%%%%%%%%%%%%%%%%%%%%%%%%%%%%%%%%%

\clearpage

%%%%%%%%%%%%%%%%%%%%%%%%%%%%%%%%%%%%%%%%%
%References
%%%%%%%%%%%%%%%%%%%%%%%%%%%%%%%%%%%%%%%%%

\bibliography{references}

\end{document}